\documentclass[cits]{PoS}\usepackage{amsmath,cite,color}

\title{Light Pseudoscalar Mesons: an Inverse Instantaneous
Bethe--Salpeter Glimpse}\ShortTitle{Light Pseudoscalar Mesons: an
Inverse Instantaneous Bethe--Salpeter Glimpse}
\author{\speaker{Wolfgang Lucha}\\Institute for High Energy Physics,
Austrian Academy of Sciences, Nikolsdorfergasse 18, A-1050 Vienna,
Austria\\E-mail: \email{Wolfgang.Lucha@oeaw.ac.at}}\author{Franz
F.~Sch\"oberl\\Faculty of Physics, University of Vienna,
Boltzmanngasse 5, A-1090 Vienna, Austria\\E-mail:
\email{franz.schoeberl@univie.ac.at}} 

\abstract{Achieving self-consistent simultaneous interpretations
of pions and kaons as bound states of quark and antiquark and as
the (almost) massless boson states related, according to
Goldstone's theorem, to the dynamical, and explicit, breakdown of
the chiral symmetries of QCD still represents a major challenge.
Applying inversion techniques to conveniently simplified versions
of the homogeneous Bethe--Salpeter equation, governing bound
states in quantum field theory, enables us to get along~a
straightforward route a qualitative idea of how the underlying
effective interaction might look like.}

\FullConference{The European Physical Society Conference on High
Energy Physics\\5--12 July 2017\\Venice, Italy}

\begin{document}
For ground-state pseudoscalar mesons, their description as
quark--antiquark bound states ought to reflect also their (near)
masslessness, demanded by Goldstone's theorem for bosons related
to~the dynamical (and explicit) breaking of the chiral symmetries
of quantum chromodynamics, the theory of the strong interactions.
The underlying \emph{effective\/} interactions enabling such
combined picture can be elucidated or explored by inversion
\cite{WL13,WLPoS} of the Bethe--Salpeter formalism, suitably
simplified by allowing for adequate instantaneous, hence
three-dimensional, reductions \cite{WL05:LS,WL05:DQ}. From the
latter,~we may extract information \cite{WL15,WL16,WL16n} in form
of configuration-space central potentials \cite{WL16i,WL16@,WL16c}
$V(r)$, $r\equiv|\mathbf{x}|$.

Strictly respecting \emph{Poincar\'e covariance\/}, the
homogeneous Bethe--Salpeter equation governing
fermion--antifermion bound states is constructed from an integral
kernel that subsumes the effective interactions responsible for
the formation of the bound states and the propagators of the
constituents of the bound states. Its solutions capture the
distribution of the relative momenta of the constituents.

In view of evident difficulties to deduce the interaction kernel
from quantum chromodynamics, we suggest to approach some of the
information assumed to be encoded therein by applying more or less
standard inversion procedures to sophisticated simplifications of
the Bethe--Salpeter formalism, cast, for reasonably trivial
dependence of kernel \cite{SE} and propagators \cite{WL05:LS} on
the time components of all relevant momenta and flavour-, Fierz-
and spherically symmetric interactions, into a bound-state problem
for the Bethe--Salpeter solution's radial components. Only
\emph{one\/} of the latter, denoted~by $\varphi_2$, matters for
\emph{pseudoscalar\/} mesons. Knowledge of solutions then sheds
light on the basic interactions.

Invariance of a quantum field theory under chiral transformations
entails identities which relate this theory's Green functions. In
the chiral limit, one such identity connects \cite{MRT,WL15} quark
propagator and Bethe--Salpeter solution for a zero-mass
pseudoscalar meson in its center-of-momentum~frame.

\begin{figure}[hbt]\begin{center}
\includegraphics[scale=1.65358]{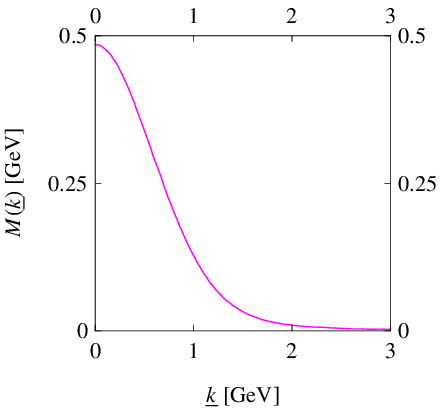}\qquad
\includegraphics[scale=1.65358]{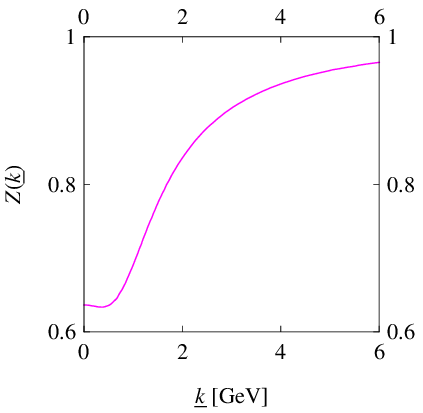}
\caption{Euclidean quark propagator functions in chiral limit:
$M(\underline{k})$ (left) and $Z(\underline{k})$ (right) vs.\
$\underline{k}\equiv\sqrt{\underline{k}^2}$ \cite{PM00}.}
\label{Fig:QPF}\end{center}\end{figure}

Owing to Poincar\'e covariance, just \emph{two\/} Lorentz-scalar
functions characterize the propagator of a quark (of four-momentum
$p$), its mass $M(p^2)$ and its wave-function renormalization
$Z(p^2)$,~which we take from a solution \cite{PM99} of this quark
propagator's \underline{Euclidean-space} equation of motion
(Fig.~\ref{Fig:QPF}).

\begin{figure}[hbt]\begin{center}
\includegraphics[scale=1.56453]{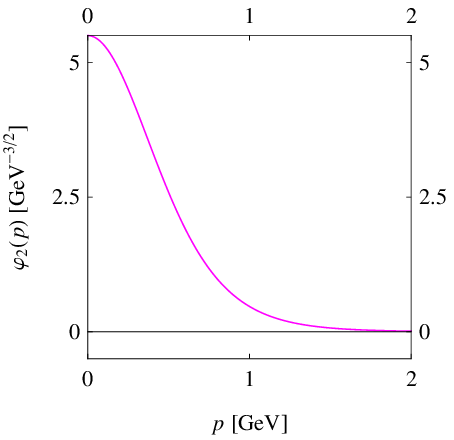}\qquad
\includegraphics[scale=1.56453]{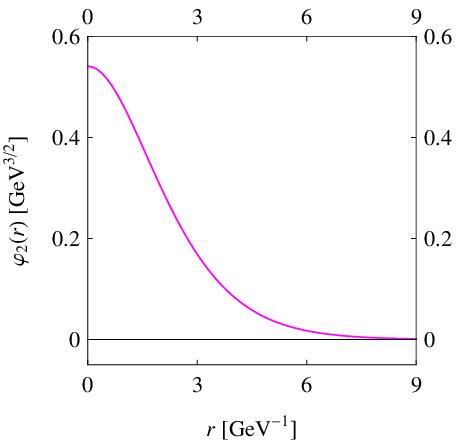}
\caption{Nonzero Salpeter component in momentum [$\varphi_2(p)$,
left] and configuration [$\varphi_2(r)$, right] space
\cite{WL16i}.}\label{Fig:ISC}\end{center}\end{figure}

Inserting the propagator functions' behaviour plotted in
Fig.~\ref{Fig:QPF} into the aforementioned identity opens the path
\cite{WL16i} to the ground-state solution $\varphi_2(p)$, with
radial variable $p\equiv|\mathbf{p}|$, and $\varphi_2(r)$
(Fig.~\ref{Fig:ISC}).\pagebreak

\begin{figure}[hbt]\begin{center}
\includegraphics[scale=1.455543]{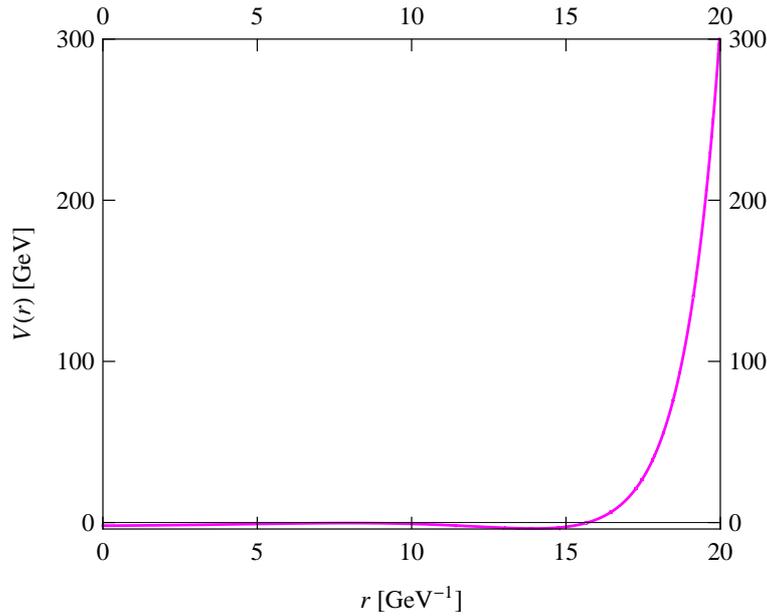}
\caption{Configuration-space central potential $V(r)$
\cite{WL16i}, arising from a chiral-limit quark propagator
\cite{PM99,PM00} as input by inverting our three-dimensional
bound-state equation \cite{WL05:LS} with Fierz-symmetric
interaction kernel.} \label{Fig:PoV}\end{center}\end{figure}

Our bound-state equation in configuration space then yields the
potential $V(r).$ This potential's unexpected square-well shape
\emph{resemblance\/} we consider as this exercise's true
\emph{quintessence\/} (Fig.~\ref{Fig:PoV}).

Not surprisingly, a brief scrutiny with the naked eye reveals
\cite{WL16i} that the ground-state solution of our
three-dimensional bound-state equation with potential $V(r)$ fixed
by inversion \emph{inevitably\/} entails a reasonable size of the
pion: its average interquark distance, $\langle
r\rangle=0.483\;\mbox{fm},$ and root-mean-square radius,
$\sqrt{\langle r^2\rangle}=0.535\;\mbox{fm},$ predicted by the
starting point $\varphi_2(r)$ of our inversion approach match the
experimentally determined value of its electromagnetic charge
radius, $\sqrt{\langle
r_\pi^2\rangle}=(0.672\pm0.008)\;\mbox{fm}$.

\begin{figure}[hbt]\begin{center}
\includegraphics[scale=1.56453]{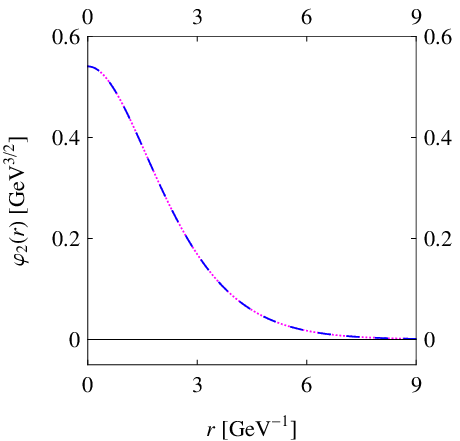}\qquad
\includegraphics[scale=1.56453]{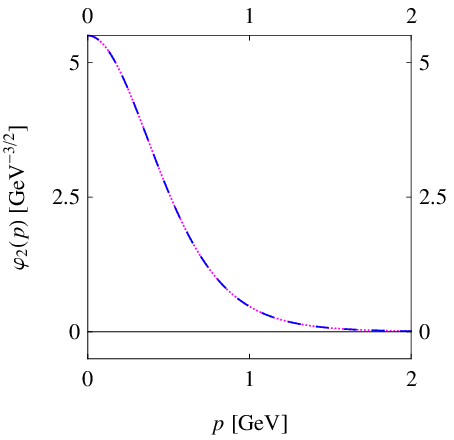}
\caption{Actual ground-state solutions
\textcolor{magenta}{(dotted)} to our three-dimensional bound-state
equation \cite{WL05:LS}, obtained straightforwardly by application
of variational techniques [$\varphi_2(r)$, left] or a conversion
to an equivalent matrix eigenvalue problem [$\varphi_2(p)$,
right], vs.\ the initial Salpeter component
\textcolor{blue}{(dashed)}, this inversion's \emph{starting
point\/}.}\label{Fig:GSS}\end{center}\end{figure}

Consistency of our inversion results can be established by
numerically solving, for the effective interaction potential
derived thereby, the bound-state equation either variationally
(Fig.~\ref{Fig:GSS}, left) or by expansion over a suitable basis
of function space (Fig.~\ref{Fig:GSS}, right). Our approach
passes, of course, this test with flying colours. In both cases,
the overlap of wave-function input and outcome equals~unity.

\begin{thebibliography}{99}
\bibitem{WL13}W.~Lucha and F.~F.~Sch\"oberl, Phys.~Rev.~D {\bf 87}
(2013) 016009, arXiv:1211.4716 [hep-ph].
\bibitem{WLPoS}W.~Lucha, Proc.~Sci., EPS-HEP 2013 (2013) 007,
arXiv:1308.3130 [hep-ph].
\bibitem{WL05:LS}W.~Lucha and F.~F.~Sch\"oberl, J.~Phys.~G {\bf 31}
(2005) 1133, arXiv:hep-th/0507281.
\bibitem{WL05:DQ}Z.-F.~Li, W.~Lucha, and F.~F.~Sch\"oberl,
Mod.~Phys.~Lett.~A {\bf 21} (2006) 1657, arXiv:hep-ph/0510372.
\bibitem{WL15}W.~Lucha and F.~F.~Sch\"oberl, Phys.~Rev.~D {\bf 92}
(2015) 076005, arXiv:1508.02951 [hep-ph].
\bibitem{WL16}W.~Lucha and F.~F.~Sch\"oberl, Phys.~Rev.~D {\bf 93}
(2016) 056006, arXiv:1602.02356 [hep-ph].
\bibitem{WL16n}W.~Lucha and F.~F.~Sch\"oberl, Phys.~Rev.~D {\bf 93}
(2016) 096005, arXiv:1603.08745 [hep-ph].
\bibitem{WL16i}W.~Lucha and F.~F.~Sch\"oberl, Int.~J.~Mod.~Phys.~A
{\bf 31} (2016) 1650202, arXiv:1606.04781 [hep-ph].
\bibitem{WL16@}W.~Lucha, EPJ Web Conf.~{\bf 129} (2016) 00047,
arXiv:1607.02426 [hep-ph].
\bibitem{WL16c}W.~Lucha, EPJ Web Conf.~{\bf 137} (2017) 13009,
arXiv:1609.01474 [hep-ph].
\bibitem{SE}E.~E.~Salpeter, Phys.~Rev.~{\bf 87} (1952) 328.
\bibitem{MRT}P.~Maris, C.~D.~Roberts, and P.~C.~Tandy,
Phys.~Lett.~B {\bf 420} (1998) 267, arXiv:nucl-th/9707003.
\bibitem{PM99}P.~Maris and P.~C.~Tandy, Phys.~Rev.~C {\bf 60}
(1999) 055214, arXiv:nucl-th/9905056.
\bibitem{PM00}P.~Maris, in \emph{Proc.~Int.~Conf.~on Quark
Confinement and the Hadron Spectrum IV\/}, eds.~W.~Lucha and
K.~Maung Maung (World Scientific, Singapore, 2002), p.~163,
arXiv:nucl-th/0009064.\end{thebibliography}
\end{document}